# Implementing Microwave Impedance Microscopy in a Dilution Refrigerator

Zhanzhi Jiang[1], Su Kong Chong[2], Peng Zhang[2], Peng Deng[2], Shizai Chu[1], Shahin Jahanbani[1], Kang L. Wang[2], Keji Lai[1, a]

[1]Department of Physics, University of Texas at Austin, Austin, TX 78712

[2]Department of Electrical and Computer Engineering, University of California, Los Angeles, CA 90095

[a]The author to whom correspondence may be addressed: kejilai@physics.utexas.edu

## Abstract

We report the implementation of a dilution-refrigerator-based scanning microwave impedance microscope (MIM) with a base temperature of ~ 100 mK. The vibration noise of our apparatus with tuning-fork feedback control is as low as 1 nm. Using this setup, we have demonstrated the imaging of quantum anomalous Hall states in magnetically (Cr and V) doped $(Bi, Sb)_2Te_3$ thin films grown on mica substrates. Both the conductive edge modes and topological phase transitions near coercive fields of Cr-doped and V-doped layers are visualized in the field-dependent results. Our work establishes the experimental platform for investigating nanoscale quantum phenomena under ultralow temperatures.

# I. INTRODUCTION

Near-field microwave microscopy is an actively evolving research field in the past few decades. A wide variety of near-field probes and microwave electronics have been proposed and implemented to spatially resolve the local permittivity, permeability, conductivity, and nonlinear response of functional materials and devices.[1,2] For the study of quantum materials, the technique is usually configured in cryogenic environments such that the thermal energy is comparable to or smaller than intrinsic quantum mechanical energy scales in the system. For instance, cryogenic microwave microscopy has been demonstrated to investigate two-level quantum systems,[3-5] high-Tc superconductors,[6-10] and degenerately doped semiconductors.[11] Continued effort in this field will bring in much knowledge on advanced quantum systems at a length scale far below the freespace wavelength of microwave radiation.

A specific implementation of near-field microwave microscopy, the microwave impedance microscopy (MIM),[12] detects the small change of tip-sample impedance at gigahertz (GHz) frequencies, from which the nanoscale permittivity and conductivity of materials can be spatially resolved. Owing to the high spatial resolution, exquisite electrical sensitivity, and sub-surface imaging capability, the technique has found widespread applications in quantum material research.[13,14] In the past decade, the development of cryo-MIM setups has enabled the investigation of mesoscopic quantum physics such as correlated electronic states in complex oxides[15-19] and moiré superlattices,[20-23] as well as topological edge states in semiconductor quantum wells[24-26] and exfoliated atomically-thin layered materials.[27-31] To date, most cryo-MIM experiments have been performed in helium-4 (He4) cryostats with a base temperature ($T$) of 2 – 4 K.[11,15-31] The only sub-Kelvin MIM work was carried out in a He3 refrigerator with a limited hold time and a base $T$ of ~ 450 mK.[32] In contrast, a He3/He4 dilution refrigerator (DR) will enable continuous operation at 100 mK or lower temperatures. The implementation of a DR-based MIM is therefore desirable for probing local electronic properties in ultralow-energy quantum physics, such as fractional quantum Hall effect (FQHE)[33] in semiconductor heterostructures and quantum anomalous Hall effect (QAHE)[34] in magnetically doped topological insulators.

In this paper, we report the design and construction of a cryo-MIM with tuning-fork feedback control[35,36] on a dilution refrigerator. The vibration isolation of the system is capable of keeping the topographic noise as low as ~ 1 nm. We demonstrate the visualization of QAHE states in magnetically (Cr and V) doped (Bi, Sb)$_2$Te$_3$ thin films grown on mica substrates at ~ 100 mK.

Topological phase transitions near the coercive fields of Cr and V are also observed in the field-dependent data. This work establishes the milli-Kelvin MIM platform for spatially resolving quantum phenomena under ultralow temperatures.

## II. EXPERIMENTAL SETUP

The milli-Kelvin MIM is implemented in a dilution refrigerator (MNK126-500, Leiden Cryogenics, with a nominal cooling power of 500 μW at 120 mK) inside a liquid helium dewar with an 8 T superconducting magnet. As schematically shown in Figure 1a, the cryostat is housed in a steel enclosure underneath the ground level. The entire structure rests on 4 pneumatic isolators (Micro-g Air Isolators, Technical Manufacturing Corporation) that can effectively filter out floor vibrations above 4 Hz. Two concrete boxes (not shown) are also installed to damp out vibrations from the 1 K pot and still pumps. The cryostat, gas handling system, and measurement electronics are all placed inside a metal-shielded room to reduce electrical and acoustic noise from the environment.

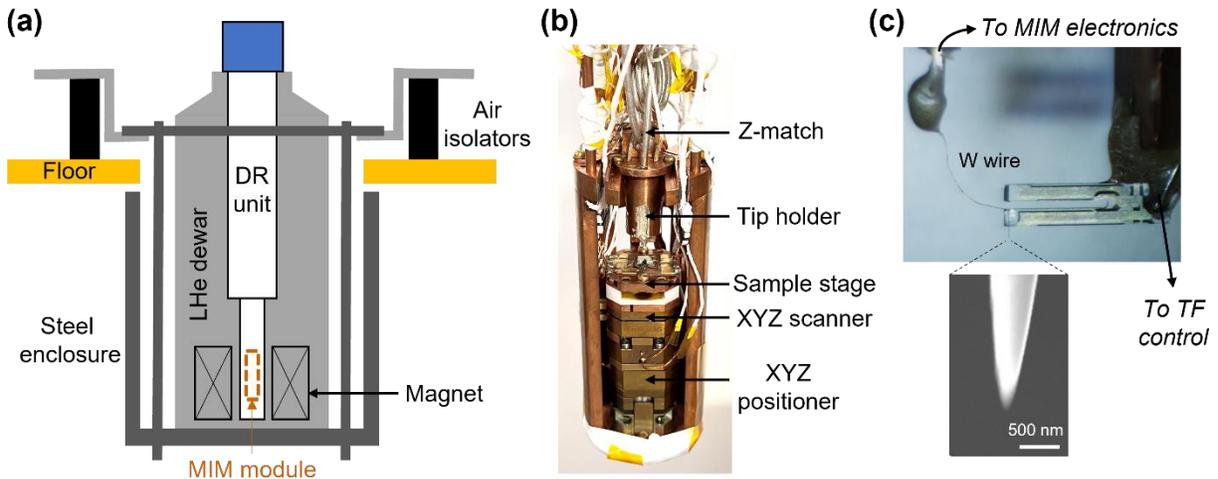

FIG. 1 (color online). (a) Schematic of the dilution refrigerator and vibration isolation. (b) Picture of the MIM module with various components. (c) Close-up view of the quartz tuning fork and a tungsten wire glued to one prong. The inset shows an SEM image of the etched tip apex.

As pictured in Figure 1b, the MIM module is mounted at the bottom of the mixing chamber of the DR. The module consists of the positioning and scanning stages, sample and tip holders, and a few microwave components (see below). A set of piezoelectric scanners (ANSxyz100, AttoCube Systems AG) and positioners (ANPx101 and ANPz101, AttoCube Systems AG) are stacked together to move the sample stage. At milli-Kelvin temperatures, the maximum scan window in a single frame is 30 μm × 30 μm, and the full positioning range is 3 mm × 3 mm. The scan speed is kept below 100 nm/s to avoid the heating effect due to the hysteresis loss and internal friction in the piezo scanner. A pair of flexible copper braids are used as the cold finger to connect the sample stage to the mixing chamber. In our current setup, the base temperatures at the mixing chamber and at the sample stage are 60 mK and 100 mK, respectively. The sample temperature is mostly limited by the heat load from the piezo stacks, which are difficult to thermalize at milli-Kelvin temperatures.[37] While the bottom Z-positioner is in good thermal contact with the copper housing, the cooling of XY-positioners and XYZ-scanners is much less efficient. In fact, if the Teflon spacer between the sample stage and piezo scanner (as seen in Figure 1b) is removed, the sample temperature will rise to ~ 500 mK. In the future, we plan to connect all piezo stacks to the copper housing with copper wires,[38] which may further reduce the sample temperature.

Figure 1c shows a close-up view of the tip assembly. A quartz tuning fork (ECS-.327-12.5-8X, Digikey Electronics) is used for controlling tip-sample distance and performing topography imaging. The magnetic leads of the tuning fork are replaced by non-magnetic CuNi wires to minimize magnetic effects during the field sweep. A tungsten wire with a diameter of 25 μm is etched to a sharp tip with a radius of curvature around 100 nm at the apex (scanning electron microscopy image in the inset of Figure 1c). The wire is then glued to one prong of the tuning fork by fast-setting epoxy (Devcon 14250). The extra mass due to the tungsten wire and glue breaks the symmetry between the two prongs and causes a shift in the resonant frequency of the TF from its unloaded value $\omega_0$ = 32,768 Hz. On the other hand, this effect can be compensated by strain in the tungsten wire if $\Delta k/\Delta m = \omega_0^2$, where $\Delta k$ and $\Delta m$ are the effective spring constant and the loaded mass, respectively.[39] Using the Young's module of tungsten and the mass of the epoxy and tungsten wire calculated from their density and volume, we estimate that a wire length of 1 – 2 mm will lead to the perfect compensation between mass and strain effects. In practice, however, it is difficult to handle such a short wire with hands and we are using a wire length of 4 – 5 mm for the convenience of tip gluing. With this length, the resonance frequency of the tuning fork decreases by ~ 500 Hz compared with $\omega_0$ and the estimated oscillation amplitude of its base is about 3% of that of its prongs. We have found that the tuning fork stays as a good harmonic oscillator (Q = 5,000 – 30,000) at the base temperature, which is optimal for the feedback control.

Finally, we use a tapping amplitude of 10 – 30 nm in the experiment, which is much smaller than the tip diameter (100 – 200 nm) and does not affect the spatial resolution.[35,36]

Figure 2a shows the schematic of tuning-fork control loops to perform topography imaging based on the frequency-modulation feedback mechanism.[35] We use the Nanonis BP5 SPM controller to excite the fork and regulate its response. A phase control loop (PLL) is used to keep the tuning fork on resonance and track its resonant frequency $f_{res}$. As the sample is brought into close proximity to the tip by the Z-positioner and the Z-scanner, $f_{res}$ is shifted due to the tip-sample force. After the sample surface is located, a PID loop regulates the height of the sample stage by controlling the Z-scanner voltage such that the frequency stays at the set point $f_{set}$ as the tip is scanned across the sample. The surface topography can then be mapped out by recording the Z-scanner voltage during the raster scan. A second PID loop is also used to maintain a constant tapping amplitude for quantitative MIM experiments. As long as the Z-feedback is operating properly, we do not observe obvious tip degradation over a typical experimental window of several weeks.

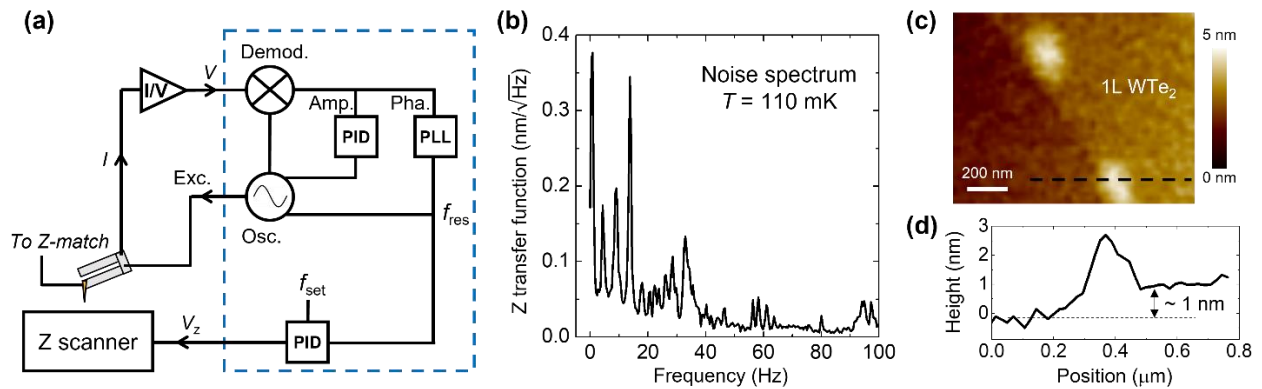

FIG. 2. (color online) (a) Schematic of the control loops for frequency-modulation-based topography imaging with the quartz tuning fork. The dashed box represents the Nanonis SPM controller. (b) Transfer function of the noise spectrum in topography imaging, showing various characteristic peaks in the Z noise. (c) Topographic image of a monolayer WTe$_2$ sample encapsulated by hBN flakes. (d) Profile along the dashed line in (c), showing the height of the monolayer as ~ 1 nm.

The noise spectrum of the topography imaging can be obtained by holding the tip at a fixed point on the sample surface and measuring the Z transfer function (Fourier transformation of time-domain signals). As shown in Figure 2b, when all pumps are in operation and the sample is at the base temperature, dominant features in the noise spectrum are all below 40 Hz, as expected for the typical building vibration. Moreover, all peaks are smaller than 400 pm/√Hz, indicative of good vibration isolation in our setup. As an example of the tuning-fork performance, Figure 2c shows the topographic image of a monolayer WTe$_2$ sample encapsulated by top and bottom hBN flakes taken at 110 mK. The line profile in Figure 2d indicates that the instrument can resolve the heights of a monolayer (~ 1 nm) and interface bubble (~ 2 nm). The noise level of the DR-based MIM under working conditions is thus below 1 nm.

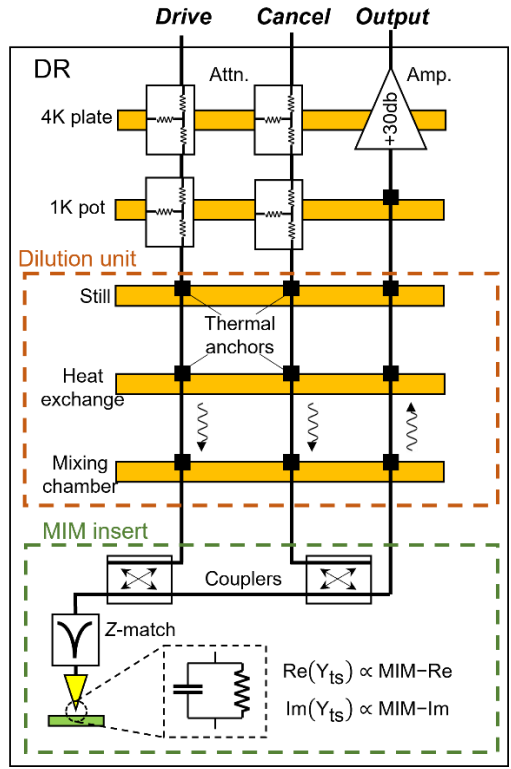

FIG. 3. (color online) Schematic of various microwave components of the MIM at different stages inside the dilution refrigerator.

Figure 3 shows the configuration of microwave components inside the DR unit. Three coaxial cables– *Drive*, *Cancel*, and *Output* – are installed to transmit microwave signals between the DR top flange and the MIM insert. The *Drive* line sends the excitation signal to the tip through the Z-match section. The *Cancel* line delivers a microwave signal whose amplitude and phase are adjusted to eliminate the common-mode signal before the scanning. The signal from the tip is pre-amplified by a 30 dB cryogenic low-noise amplifier (CITLF4, Cosmic Microwave Technology)[24] and sent to the *Output* line, after which the signal is further amplified and demodulated by external microwave electronics. The two output channels are proportional to the imaginary (MIM-Im) and real (MIM-Re) parts of the tip-sample admittance.[12] Note that for tapping-mode MIM, a second demodulation at the resonant frequency of the tuning fork is needed to yield the final outputs.[35,36] Details of the MIM electronics and dual demodulation can be found in Ref. [36].

In order to balance the heat load and power transmission, we use multiple types of semi-rigid coaxial cables to construct the microwave signal lines. The section between the DR top flange and the 4K plate is made of 047 CuNi coaxial cables.[37] Between the 4K plate and the mixing chamber, stainless steel cables with silver plating on the center conductor (SSS-SS 047, IntelliConnect) are used. In contrast, good thermal conductivity is desired between the mixing chamber and the tip holder, thus the usage of copper coaxial cables. On each cold plate, we install thermal anchors to thermalize the outer conductors. The cables are terminated with gold-plated CuBe or brass SMA connectors mounted on OFHC plates, which are screwed onto the cold plate. On the *Drive* and *Cancel* lines, a set of RF attenuators are installed at the 4K and 1K-pot plates to further thermalize the center conductor. According to the literature, an attenuation of 10 – 20 dB is needed to minimize the radiative noise from the high-$T$ side to the low-$T$ side of each cable.[40] In our application, however, we usually need to perform MIM measurements at various stages (room temperature, liquid nitrogen, liquid helium, and base temperature) of the cool-down process to check the tip and sample conditions. As a result, we use a modest attenuation of 3 – 6 dB to achieve a good balance between noise reduction and power consideration for our MIM experiment.

As shown in Figure 3, the MIM insert is mounted at the bottom of the mixing chamber. Here the microwave signals are routed to the *Drive*/*Cancel*/*Output* lines by two broadband (0.5 – 18 GHz) directional couplers. To ensure efficient power transfer to the tip, the excitation signal is guided to an impedance match (Z-match) section for the normal 50 $\Omega$ transmission lines. Following the design in Ref. [35], we use a 20 cm hand-formable cable with a 0.1 pF series capacitor for impedance match. Since it is desirable to apply a tip bias in many MIM experiments, we also place a 10 M$\Omega$ chip resistor in parallel with the capacitor such that the tip can be held at a well-defined DC potential. This configuration gives a fundamental Z-match frequency of ~ 0.5 GHz and higher-order resonances at integer multiples of this frequency for broadband studies. For the proof-of-principle experiment below, we operate at around 2.8 GHz with a return loss of ~ 20 dB at the base temperature. The sensitivity of admittance (inverse impedance) detection of the MIM electronics is ~ 1 nS at this frequency.[12]

**III. RESULTS AND DISCUSSIONS**

We demonstrate the milli-Kelvin MIM experiment by imaging a magnetically doped topological insulator (Bi, Sb)$_2$Te$_3$ (BST) sample grown on mica by molecular beam epitaxy (MBE). As illustrated in Figure 4a, the active material consists of 3 quintuple layers (QLs) of V-doped BST, 6 QLs of undoped BST, and 3 QLs of Cr-doped BST.[41] The heterostructure, together with a thin (~ 20 nm) mica layer, is exfoliated from the substrate and transferred to a heavily-doped silicon wafer with 285 nm SiO$_2$ on the top. The device is patterned into a standard Hall bar with Cr/Au electrodes, as shown in Figure 4b. Since there is no optical access to the tip-sample assembly at low temperatures, we also pattern T-shaped markers on the device in order to guide the tip towards the Hall bar.

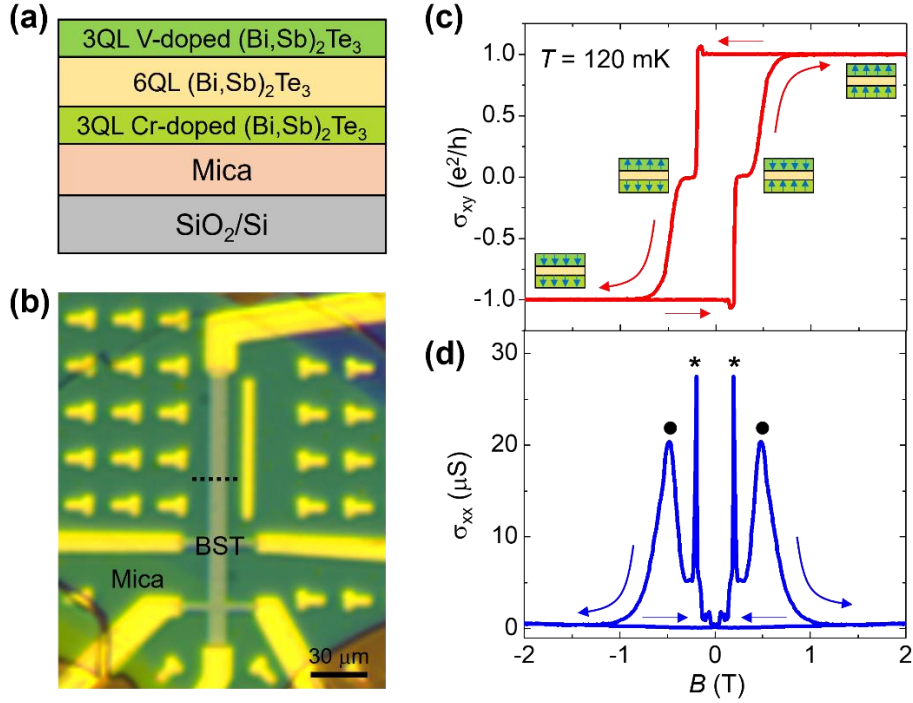

FIG. 4. (color online) (a) Cross-sectional layer structure of the sample. (b) Optical image (top view) of the Hall-bar device. (c) Magnetic field dependence of $\sigma_{xy}$ and (d) $\sigma_{xx}$ at $T = 120$ mK and $V_g = -30$V in this sample. The field is applied perpendicular to the sample surface. Arrows indicate the field sweep direction. The asterisks and dots denote the $\sigma_{xx}$ peaks associated with topological phase transitions in the Cr-doped and V-doped BST layers, respectively.

Figures 4c and 4d show the *B*-field dependence of Hall ($\sigma_{xy}$) and longitudinal ($\sigma_{xx}$) conductivities, respectively, at 120 mK and back-gate voltage $V_g = -30$ V. Starting from a large positive field, the quantization of $\sigma_{xy}$ at $+e^2/h$ and the nearly zero $\sigma_{xx}$ for fields above 1 T are clear signatures of the C = 1 QAHE state, in which the external field aligns the magnetization of both V-doped and Cr-doped layers. As the $B$ field is swept toward the C = −1 QAHE state, two prominent $\sigma_{xx}$ peaks are observed in the transport data, accompanied by two steps of change in $\sigma_{xy}$, first to zero and then to $-e^2/h$. These features are assigned to the sequential magnetization flipping of Cr- and V-doped BST layers, whose coercive fields are substantially different.[41] In between the two $\sigma_{xx}$ peaks, the minimum in $\sigma_{xx}$ and the zero $\sigma_{xy}$ plateau indicate the formation of an axion insulator state, where the magnetization of the two doped layers is antiparallel with each other.[41] The same features occur at opposite fields as the $B$-field is swept from –2 T to 2 T. The sweeping rate is kept at 2 mT/s such that the sample temperature is stable. We notice that neither the QAHE state nor the axion insulator state is perfect in this sample due to the presence of residual bulk carriers at $V_g$ = −30 V (higher $V_g$ leads to gate leakage). For the former, $\sigma_{xx}$ continues to rise as increasing fields deep into the QAHE state. For the latter, the minimum in $\sigma_{xx}$ does not reach zero. Further improvement in the growth procedure is therefore needed for a better quality of the sample.

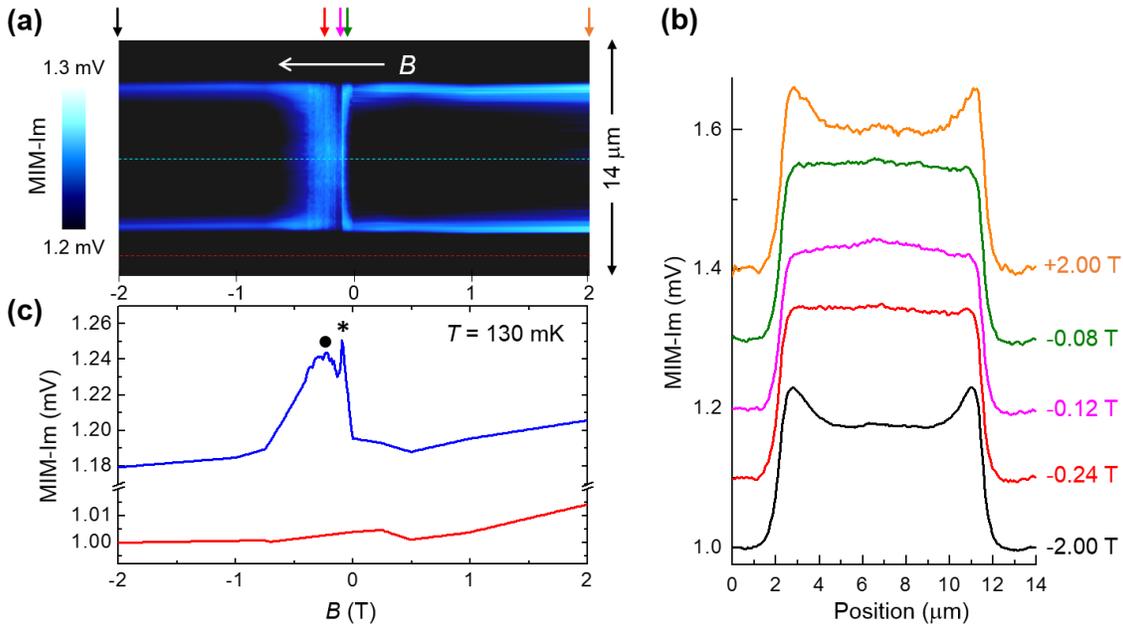

FIG. 5. (color online) (a) MIM-Im line profile along the dashed line in Figure 4b as the magnetic field is swept from 2T to −2T at 130 mK. Note that the apparent asymmetry between positive and negative fields is due to a slow background drift in the signal. (b) Selected profiles at 5 representative fields, as indicated by arrows in (a). The curve taken at – 2.00 T corresponds to the actual MIM data. All other line cuts are offset by 0.1 mV between adjacent curves for clarity. (c) Field-dependent MIM-Im signals at the center of the Hall bar (blue curve) and on the substrate (red curve). The asterisk and dot again denote the peaks associated with the switching of Cr-doped and V-doped layers, respectively.

The MIM results of this heterostructure device at $T = 130$ mK are shown in Figure 5. Since the MIM-Im signal is a monotonic function of the local conductivity,[12] we only present data from this channel for the proof of concept in this paper. Details of the MIM imaging and quantitative analysis will be reported elsewhere. Figure 5a displays the evolution of MIM-Im signals as the $B$-field is swept from 2 T to −2 T along a single line across the Hall bar (dashed line in Figure 4b).[32] The characteristic line profiles at five different fields are shown in Figure 5b. When the system is in the C = 1 QAHE state (2 T > $B$ > 0 T), the MIM-Im profile exhibits two prominent peaks near the physical boundaries, consistent with the existence of chiral edge modes under this condition. Around $B = -0.08$ T, the edge state disappears, and the sample is uniformly conductive. The MIM signal in the bulk drops slightly at $B = -0.12$ T and then increases again around $B = -0.24$ T. Beyond −1 T, the edge state reappears, as seen in the $B = -2$ T (C = −1 QAHE) data. The same evolution can be observed by plotting the field-dependent MIM-Im signals through the center of the sample in Figure 5c. Here the narrow and broad peaks correspond to the magnetization flipping of Cr- and V-doped BST layers, respectively, and the dip between them represents the axion insulator state. The results resemble those in the $\sigma_{xx}$ vs $B$ curve obtained by the transport measurement. Note that there appears to be a background drift of the MIM signals both on the Hall bar and on the substrate (Figure 5c) over this long experiment (> 24 hours), whose origin is subject to future studies.

Finally, we briefly discuss the milli-Kelvin MIM results on this sample. By comparing Figures 4d and 5c, one can see that the topological phase transitions occur at different fields between the transport (−0.19 T for Cr-BST and −0.48 T for V-BST) and MIM (−0.08 T for Cr-BST and −0.24 T for V-BST) data. This is presumably due to the fact that the transport experiment is performed at a much faster pace (< 2 hours from 2 T to −2 T) than the MIM imaging (> 24 hours from 2 T to −2 T). As a result, the phase-transition fields in the MIM measurement are closer to the actual coercive fields of the magnetically doped BST layers. In addition, the apparent edge width of 0.6 ~ 1 μm is likely due to the finite tip size and residual bulk conductance. Future experiments using sharper tips and better samples are expected to provide more accurate information on the dimension of the edge states.[42,43] Improved spatial resolution may also allow us to visualize the internal conductive edge modes along magnetic domain walls during the topological phase transitions.[44]

## IV. CONCLUSIONS

In summary, we have successfully implemented a scanning microwave impedance microscopy in the dilution refrigerator platform with an operating temperature of ~ 100 mK. The vibration isolation in our system ensures that the topographic noise of the tuning-fork-based sensor is below 1 nm. Using this setup, we have demonstrated the milli-Kelvin MIM imaging of quantum anomalous Hall and axion insulator states in magnetically doped $(Bi, Sb)_2Te_3$ thin films. The topological phase transitions associated with magnetization switching of the Cr- and V-doped layers are seen in both transport and MIM experiments. Our work opens up new avenues for research on quantum materials in the ultralow temperature regime.

## ACKNOWLEDGMENTS

This research was funded by the U.S. Department of Energy (DOE), Office of Science, Basic Energy Sciences, under Award No. DE-SC0019025. The instrumentation work was supported by the Army Research Office and accomplished under Grants No. W911NF-17-1-0542 and W911NF-18-1-0467. K.L.W. acknowledges support from the National Science Foundation the Quantum Leap Big Idea under Grant No. 1936383 and the U.S. Army Research Office MURI program under Grant No. W911NF-20-2-0166. The authors thank P. Wang and S. Wu for providing the $WTe_2$ sample and Z. Yao for sharing the equipment.

## DATA AVAILABILITY

The data that support the findings of this study are available from the corresponding author upon reasonable request.